\documentclass[%
 reprint,
superscriptaddress,
 amsmath,amssymb,
 aps,
]{revtex4-2}

\usepackage{graphicx}% Include figure files
\usepackage{dcolumn}% Align table columns on decimal point
\usepackage{comment}
\usepackage{bm}% bold math
\usepackage{color}
\DeclareMathOperator{\Tr}{Tr}
\usepackage{ulem}

\newcommand{\dd}{\mathrm{d}}

\begin{document}

\title{Enhancing (quasi-)long-range order in a two-dimensional driven crystal  }

\author{R. Maire}
\affiliation{Universit\'e Paris-Saclay, CNRS, Laboratoire de Physique des Solides, 91405 Orsay, France}
\author{A. Plati}
\email{andrea.plati@universite-paris-saclay.fr}
\affiliation{Universit\'e Paris-Saclay, CNRS, Laboratoire de Physique des Solides, 91405 Orsay, France}
%\author{ F. Smallenburg}
%\affiliation{Universit\'e Paris-Saclay, CNRS, Laboratoire de Physique des Solides, 91405 Orsay, France}
%\author{G. Foffi}
%\affiliation{Universit\'e Paris-Saclay, CNRS, Laboratoire de Physique des Solides, 91405 Orsay, France}
\date{\today}

\begin{abstract}
It has been recently shown that 2D systems can exhibit crystalline phases with long-range translational order showcasing a striking violation of the Hohenberg-Mermin-Wagner (HMW) theorem which is valid at equilibrium. 
This is made possible by athermal driving mechanisms that inject energy into the system without exciting long wavelength modes of the density field, thereby inducing hyperuniformity.
However, as thermal fluctuations are superimposed on the non-equilibrium driving, long-range translational order is inevitably lost. Here, we discuss the possibility of exploiting non-equilibrium effects 
to suppress arbitrarily large density fluctuations even when a global thermal bath is coupled to the system. We introduce a model of a harmonic crystal driven both by a global thermal bath and by a momentum conserving noise, where the typical observables related to density fluctuations and long-range translational order can be analytically derived and put in relation. This model allows us to rationalize the violation of the HMW theorem observed in previous studies through the prediction of large-wavelength phonons which thermalize at a vanishing effective temperature when the global bath is switched off. The conceptual framework introduced through this theory is then applied to numerical simulations of a hard-disk solid in contact with a thermal bath and driven out-of-equilibrium by active collisions. Our numerical analysis demonstrates how varying driving and dissipative parameters can lead to an arbitrary enhancement of the quasi-long-range order in the system regardless of the applied global noise amplitude. Finally, we outline a possible experimental procedure to apply our results to a realistic granular system.
\end{abstract}
\maketitle

%\tableofcontents

\section{\label{sec:Intro} Introduction}

The celebrated Hohenberg-Mermin-Wagner (HMW) theorem \cite{hohenberg1967existence, mermin1966absence, halperin2019hohenberg} is a cornerstone of equilibrium statistical mechanics. It establishes the impossibility of obtaining long-range order through continuous symmetry breaking in 1D and 2D equilibrium spin systems with short-range interactions at finite temperature. Similarly, 2D crystals with short-range interactions exhibit translational quasi-long-range order with a correlation function decaying as a power law due to strong phonons excitations at arbitrarily large length scales \cite{mermin1968crystalline}. However, their bond-orientational order is long-range \cite{alder1957phase, frohlich1981absence, nelson2012bond, gasser2009crystallization}.

While the HMW theorem arises from equilibrium statistical physics, the majority of systems encountered in nature operate out-of-equilibrium. Consequently, the applicability of the theorem to such systems is not assured.
As an example,  a notable breakdown is observed in the flocking behavior of birds \cite{cavagna2010scale}. This process can be modeled by an active $XY$ model -- the Vicsek model \cite{vicsek1995novel}, showcasing spin-spin long-range order in 2D \cite{toner1995long, toner1998flocks, toner2005hydrodynamics, bertin2009hydrodynamic, weber2013long}, in clear violation of the HMW theorem.
% We remark here that it has been recently shown that flocking can be easily destabilized by the nucleation of droplets moving in the direction opposite to that of the ordered phase.
Various studies have further explored potential violations of this theorem in non-equilibrium spin-like systems due to forcing at multiple temperatures \cite{bassler1995existence, reichl2010phase}; advection of the order parameter through a shear flow \cite{giomi2022long,nakano2021long, minami2022origin, minami2021rainbow, ikeda2024advection}; or a colored noise \cite{ikeda2023does, Ikeda2023CorrelatedNA}. These ingredients have been shown to be crucial in order to obtain spin-spin long-range order which, in turn, can be easily destabilized by other out-of-equilibrium effects \cite{Benvegnen2023}. %However, the situation was not as clear for crystals in 2D.
However, until recently, the prerequisites for translational long-range order in 2D crystals remained unknown. Indeed, while exceptions were found in nematic crystals \cite{PhysRevLett.123.238001} or chiral active matter \cite{lei2019nonequilibrium}, Mermin-Wagner fluctuations are usually found to be enhanced in the presence of active forces \cite{PhysRevE.104.064605, PhysRevLett.131.108301, dey2024enhanced} and as a result, an increase of the lower critical dimension is often observed due to giant number fluctuations \cite{ramaswamy2003active, narayan2007long}. 
%It is important to note that, even when equilibspin
%We stress that even when  in presence of fl
%Thus, the breakdown of the HMW in non-equilibrium crystals remained the exception rather than the rule. 

However, Galliano \textit{et al.} presented compelling evidence of the breakdown of the HMW theorem in 2D crystalline systems \cite{PhysRevLett.131.047101} formed in the active state of a random organization model. The authors showed that, unlike in short-ranged equilibrium systems, a simple non-equilibrium hyperuniform crystal exhibits translational long-range order, as already hinted in Refs. \onlinecite{torquato2018hyperuniform, kim2018effect}, due to the absence of thermal fluctuations at large length scale.

Given these premises, it is important to point out that the recent observations of crystalline phases with long-range translational order in two dimensions only concern models where thermal-like fluctuations are absent by construction \cite{PhysRevLett.131.047101, kuroda2024longrange, kuroda2023microscopic, lei2019nonequilibrium}. 
These studies leave open the following key question: how is the translational order affected when thermal fluctuations cannot be neglected?

To tackle this question, in this article, we will investigate the limits of the HMW theorem and its potential breakdown in non-equilibrium systems through a theoretical model of a harmonic crystal coupled to a local bath conserving the center of mass (COM) and a global thermal-like bath. We will show that for this system, the decay of the translational long-range order is controlled by the temperature of the COM and not by the overall kinetic energy, as expected at equilibrium \cite{mermin1968crystalline}.
This theory includes the case of true long-range translational order as a particular limit and allows quantifying deviations from it due to thermal effects.
  
We next apply our theory to the time-continuous analogue \cite{maire_interplay_nodate} of a discrete random organization model in which thermal fluctuations are taken into account. It consists of a hard-disk solid in contact with a thermal bath and driven
out-of-equilibrium by active collisions. Through numerical simulations, we show that, in the absence of thermal fluctuations, this system exhibits a hyperuniform long-range ordered 2D crystal in agreement with the phenomenology observed in \cite{PhysRevLett.131.047101,kuroda2024longrange}. 
However, when thermal-like motion is taken into account, perfect long-range order is lost. Despite this,  we demonstrate how exploiting the non-equilibrium properties of the model enables us to suppress arbitrarily large density fluctuations.
This makes it possible to enhance the quasi-long-range translational order in a non-equilibrium crystal without the need to neglect or fine-tune thermal fluctuations.
Our model will also serve as a coarse-grained representation of a confined quasi-2D vibrated granular system \cite{maire_interplay_nodate} thus naturally providing a platform for investigating the enhancement of quasi-long-range translational order in experimental systems. 

The paper is organized as follows: In Sec. \ref{sec: theory} we introduce the theoretical model for a harmonic crystal and derive the analytical expression of the typical observables related to density fluctuations and long-range translational order. In Sec. \ref{sec:granular}, we report the numerical results for the non-equilibrium hard-disk solids. Finally,  Sec. \ref{sec: expe} contains the conclusion and a brief discussion about a possible experimental procedure to enhance quasi-long-range order in a realistic granular system.

\section{\label{sec: theory} Theory}
\subsection{The model}

We study a  2D crystal made of $N$ particles with masses $m$ arranged on a periodic lattice of size $L\times L$ and lattice spacing $a$. Each particle on the lattice site $\bm n$ interacts with its neighbors $\{\bar{\bm n}\}$ via a harmonic interaction $K$. Additionally, every particle is coupled to two baths: a global one and a local one which conserves the momentum and COM. In order to have a well-defined equilibrium limit, both of them separately respect the fluctuation dissipation theorem (FDT) \cite{kubo1966fluctuation, marconi2008fluctuation,villamaina2009fluctuation, puglisi2012structure}. From a physical standpoint, the global bath can emerge from the coarse-graining of an external energy source such as in the case of colloids diffusing in a fluid or beads vibrating on a rough surface, while the local bath is more likely to come from an internal source such as collisions between particles, as with the noise current in fluctuating hydrodynamics \cite{landau1980course, van1999randomly, gradenigo2011fluctuating, marconi2021hydrodynamics}. The displacement $\bm u_n$ of each particle at lattice site $\bm n$ with respect to their ideal lattice position is described by the following Langevin equation:

\begin{equation}
        m\ddot{\bm{u}}_{n} =  - K\sum_{\{\bar{n}\}}( {\bm u}_{n} - \bm{u}_{\bar{n}}) + \bm{F}_{com} + \bm{F}_{loc}
    \label{eq: lattice}
\end{equation}

The first term on the right-hand side of the equation represents the harmonic interaction between nearest neighbors.

The second term $\bm{F}_{com}$ represents the coupling between the system and a global bath, which does not conserve the position of the COM:
\begin{equation}
\bm{F}_{com} = - \gamma_{com}\dot{\bm u}_{n} + \sqrt{2\gamma_{com} T_{com}}\bm{\xi}_{n},
\end{equation}
where $\gamma_{com}$ is a global damping and $T_{com}$ is the temperature of the bath. The spatial components $\alpha$ of the white noise are Gaussian and uncorrelated, with zero average: 
\begin{equation}
\langle \xi_{n}^{\alpha}(t)\xi_{m}^{\beta}(t')\rangle = \delta(t - t')\delta_{n, m}\delta^{\alpha, \beta}~~~~~~\langle \bm \xi_n(t)\rangle = 0.
\end{equation}

Finally, the last term in Eq.~\eqref{eq: lattice} represents a second bath, at temperature $T_{loc}$, conserving the momentum and the position of the COM. To fulfill this requirement, we use a discretized version of Model B \cite{hohenberg1977theory, PhysRevX.7.021007, tauber2007field, Ikedalattice, glorioso2022breakdown, das2023friction, cui2017atomic}:
\begin{equation}
    \bm{F}_{loc} = -\gamma_{loc}\sum_{\{\bar{n}\}}( { {\dot {\bm u}}}_{n} - {\dot{\bm {u}}}_{\bar{n}})  + \sqrt{2\gamma_{loc} T_{loc}} (\bm\nabla\cdot\bm\Xi_{{n}}).
\end{equation}
The conservation of $\sum_{n}\dot {\bm u}_n$ is ensured by a discrete Laplacian of $\dot{\bm  u}$ for the damping $\gamma_{loc}$, taken to be the average over first neighbors and a discrete divergence acting on a rank 2 random tensor $\bm\Xi$.  This type of damping naturally arises in granular gases where collisions tend to align particles \cite{plati2021long}, in active matter with effective alignment \cite{carprini2020alignement,marconi2021hydrodynamics} or in Dissipative Particle Dynamics \cite{espanol2017perspective, groot1997dissipative, lowe1999alternative}. More generally, with the corresponding equilibrium noise, it plays the role of a discrete hydrodynamic viscosity and can be derived from the Mori-Zwanzig formalism \cite{castellano2023mode}. Note that if every particle moves in the same direction, neither damping nor noise is applied, in line with the idea that this local noise might arise from collisions between particles and acts locally. Our model constitutes a simplified lattice equivalent to a fluctuating hydrodynamics description of a solid \cite{mabillard_nonequilibrium_2021, hiura_microscopic_2023, szamel_slow_1993, szamel_statistical_1997, fleming_hydrodynamics_1976, martin_unified_1972, mabillard_microscopic_2020}.
The variance and the average of the rank 2 tensor $\bm \Xi_n$ are fully determined by our assumption that the FDT holds separately for both noises  \cite{puglisi2012structure, landau1992hydrodynamic}:
\begin{equation}
    \langle \Xi_{n}^{\alpha, \mu}(t)\Xi_m^{\beta, \nu}(t')\rangle =
    \delta(t' - t)\delta^{\alpha,\beta}\delta^{\mu, \nu}\delta_{n, m},
    \label{eq: 1D}
\end{equation}
We further discuss the peculiarity of the noise as well as the expression of the discrete Laplacian and divergence in Appendix~\ref{sec: AppendixNoise} following Ref. \onlinecite{Cavagna2023from}.

For simplicity, in the following, we chose to work in Fourier space by performing a semi-discrete Fourier transform (see Appendix~\ref{sec: appendixFourier}) on Eq.~\eqref{eq: lattice} \cite{caprini2023entropy, chaikin_lubensky_1995}:

\begin{equation}
    \begin{split}
         \left(-mw^2  - iw (\gamma_{com} + \gamma_{loc}\omega_k^2)+ K\omega_k^2\right) \tilde{\bm{u}}_k = &\sqrt{2\gamma_{com} T_{com}}\tilde{\bm{\xi}}_k\\ +& \sqrt{2\gamma_{loc} \omega_k^2T_{loc}} \tilde {\bm\eta}_{k},
    \end{split}
    \label{eq: FinalFourier}
\end{equation}
with $\omega_k$ the dispersion relation of the lattice divided by the natural frequency of the lattice $\sqrt{K/m}$ or equivalently, the eigenvalue of the discrete Laplacian. For instance, in a square lattice, $\omega_k^2 = 2(2 - \cos(k_xa) - \cos(k_ya))$. Both noises have unit variance and zero average, from a Fourier transform we obtain:
\begin{equation}
    \begin{split}
    \langle\tilde{\xi}_k^{\alpha}(w)\tilde{\xi}_q^{\beta}(w')\rangle = \delta(w + w')\delta_{q, -k}\delta^{\alpha, \beta}~~~~~\langle \tilde{\xi}_k^\alpha(w)\rangle = 0\\
     \langle\tilde{\eta}_k^{\alpha}(w)\tilde{\eta}_q^{\beta}(w')\rangle = \delta(w + w')\delta_{q, -k}\delta^{\alpha, \beta}~~~~~\langle \tilde{\eta}_k^\alpha(w)\rangle = 0
    \end{split}\label{eq: noisesAve}
\end{equation}
The noises are simple to generate in the reciprocal space compared to the ones in real space. Moreover, we clearly see that the local bath does not act on the COM since it vanishes at $\bm k=0$ implying that, as wanted, the motion of the COM is completely governed by the global bath.

Due to the harmonicity of the crystal, Eq.~\eqref{eq: FinalFourier} is non-interacting in $k$-space and every mode reaches an independent equilibrium-like steady state at an effective $k$-dependent temperature: 
\begin{equation}
    \tilde T_{k} = \dfrac{\gamma_{com} T_{com} + \gamma_{loc} \omega_k^2T_{loc}}{\gamma_{com} + \gamma_{loc}\omega_k^2}.
    \label{eq: Teff}
\end{equation}
Since the dispersion relation must vanishes at low-$k$:  $\omega_k^2 \sim (a \bm k)^2$ \cite{chaikin_lubensky_1995} we obtain the two limiting cases:
\begin{equation}
    \begin{split}
        \tilde T_{k\to 0} &= T_{com}\\
        \tilde T_{k\to\infty}&=T_{loc}\quad\textrm{if}\quad\gamma_{loc}\gg\gamma_{com}.
    \end{split}
    \label{eq: 7}
\end{equation}
That is, the large length scale temperature is controlled by $T_{com}$ since the energy created on small length scales by the local bath is damped by $\gamma_{com}$ on every scale. While the small length scale temperature is controlled by both the global and local bath. 
In the following, we will assume that the system is more weakly coupled to the local bath compared to the global bath: $\gamma_{loc}\gg\gamma_{com}$ which leads to $\tilde T_{k\to\infty}=T_{loc}$.

The equilibrium-like nature of our equation stems from the assumption that both baths are delta correlated and that each mode is independent on the others in $k$ space. This would not be the case in a realistic system with anharmonic terms for instance and could potentially disrupt the picture given in this section. We discuss the case of an anharmonic lattice in Appendix~\ref{sec: appendix anha} and argue that the phenomenology described here should not change qualitatively.

\subsection{Hyperuniformity}
As already noted in Ref. \onlinecite{PhysRevLett.131.047101}, a key requirement to obtain a long-range ordered 2D crystal is hyperuniformity i.e. the suppression of density fluctuations at arbitrarily large length scales. Within the context of the proposed theoretical model, we derive here the analytical expression of the structure factor $S(\bm k)$ whose vanishing low-$\bm k$ limit indicates the presence of hyperuniformity in a system \cite{torquato2018hyperuniform}.

We start by computing the static displacement function (SDF) $C_{uu}(\bm k)\equiv C_{uu}(\bm k, t =0)$:

\begin{equation}
    \begin{split}
            C_{uu}(\bm k) &= \dfrac{1}{2\pi}\int \dd w \langle \tilde{\bm u}_k(w) \cdot \tilde{\bm u}_{-k}(-w)\rangle\\
            %&=\dfrac{2}{\pi}\int dw \dfrac{\gamma_{com}T_{com} + \gamma_{loc}\omega_k^2T_{loc}}{(K\omega^2_k - mw^2)^2 + (\gamma_{com}+\gamma_{loc}\omega_k^2)^2w^2}\\
            &=\dfrac{2}{K}\dfrac{\gamma_{com}T_{com} + \gamma_{loc}\omega_k^2T_{loc}}{\omega^2_k(\gamma_{com}+\gamma_{loc}\omega_k^2)} = \dfrac{2}{K\omega_k^2}\tilde T_k.
    \end{split}
    \label{eq: phonons structure factor}
\end{equation}
where Eq.~\eqref{eq: FinalFourier} and \eqref{eq: noisesAve} have been used to calculate averages over the noise.
The above identity represents the equipartition of elastic energy at an effective temperature $\tilde T_k$ which is given by Eq.~\eqref{eq: Teff}. 
%Of course, 
The standard equipartition theorem for phonons $K \omega_k^2 C_{uu}(\bm k)/2 =  T_{loc}$ or $K \omega_k^2 C_{uu}(\bm k)/2 =  T_{com}$ is recovered in the three equilibrium limits of the model, 
%genuinely equilibrium system, 
that is either when $\gamma_{com} =0, \gamma_{loc} = 0$ or $T_{com}=T_{loc}$. In the large lengths limit where $\omega_k^2\sim (a\bm k)^2$, the SDF reads:
\begin{equation}
    \begin{split}
    C_{uu}(\bm k) =& \dfrac{2T_{com}}{K(a\bm k)^2}+\dfrac{2\gamma_{loc}(T_{loc}-T_{com})}{K\gamma_{com}}\\ &+ \mathcal{O}\left((T_{loc}-T_{com} )(a \bm k)^2\right).
    \end{split}
    \label{eq: approx CUU}
\end{equation}
and diverges as $1/(a\bm k)^2$ at equilibrium or more generally when $T_{com}\neq0$. 
%As will be made clear in the next section, it is the reason for the impossibility of translational long-range order in equilibrium short-ranged low-dimensional systems \cite{chaikin_lubensky_1995}. 
However, when $T_{com} = 0$ and $\gamma_{com}\neq0$, the phonons created locally are damped over large distances and long wavelength phonons are completely suppressed since their elastic energy $K\omega_k^2C_{uu}(\bm k)/2$ goes to 0. The same type of result can be found for the static velocity factor $mC_{\dot u\dot u}(\bm k)/2=\tilde T_k$ showing a depletion of kinetic energy on large scales.

At this point, it is illuminating to note that the long-range structure factor $S(\bm k)$ and the SDF are linked by the relation \cite{kim2018effect, torquato2018hyperuniform}:

\begin{equation}
    \begin{split}
        S(\bm k) &= \left\langle \dfrac{1}{N}\left|\sum_{\bm n} e^{i\bm k\cdot (a\bm n + \bm{u}_n) }\right|^2\right\rangle\\
        &=\left\langle \left|\bm k\cdot \tilde{\bm{u}}_k\right|^2 \right\rangle + \mathcal{O}\left(\left|\bm k\cdot \tilde{\bm{u}}_k\right|^4\right)\\
       &=|\bm k|^2C_{uu}( |\bm k|) + \mathcal{O}\left(\left|\bm k\cdot \tilde{\bm{u}}_k\right|^4\right)\\
    \end{split}
    \label{eq: structurefactor important big big}
\end{equation}
Where we used the fact that in our model the longitudinal dispersion relation is equal to the transversal ones. Otherwise, the last relation only holds for longitudinal polarization of the displacement \cite{kim2018effect}. In an equilibrium system, the value of the structure factor at $\bm k=0$ is a constant proportional to the isothermal compressibility and the temperature \cite{hansen2013theory,chen2023disordered}.
%T$, 
%the structure factor goes to a constant  the isothermal compressibility  
%$S(0)=\rho  T \chi_T$ with $\chi_T$ \cite{hansen2013theory,chen2023disordered}. 
However in our case, from Eq.~\eqref{eq: approx CUU}, it follows that:
\begin{equation}
    \begin{split}
      S(\bm k) =&\dfrac{2T_{com}}{Ka^2}+\dfrac{2\gamma_{loc}(T_{loc}-T_{com})}{\gamma_{com}}\bm k^2 \\&+ \mathcal{O}\left((T_{loc}-T_{com} )a^2\bm k^4\right). 
    \end{split}
    \label{eq: structurefactor theo}
\end{equation}
When $T_{com}= 0$ and $\gamma_{com}\neq 0$, since the SDF is finite at low-$k$, the structure factor behaves as $S(\bm k)\sim \bm k^2$, unveiling the hyperuniformity of our non-equilibrium system when the COM position or equivalently the momentum is conserved by the local noise at finite global damping. This aligns with results found in Refs. \onlinecite{lei2019hydrodynamics, Ikedalattice, hexner2017noise, hexner2015hyperuniformity, mukherjee2023anomalous,PhysRevLett.131.047101, bertrand2019nonlinear} and confirms that our model includes a suitable limit to study the emergence of long-range order in dimension lower than three.

\subsection{(Quasi-)long-range order} \label{sec: qulo}

\begin{figure}
    \includegraphics[width=0.9\columnwidth,clip=true]{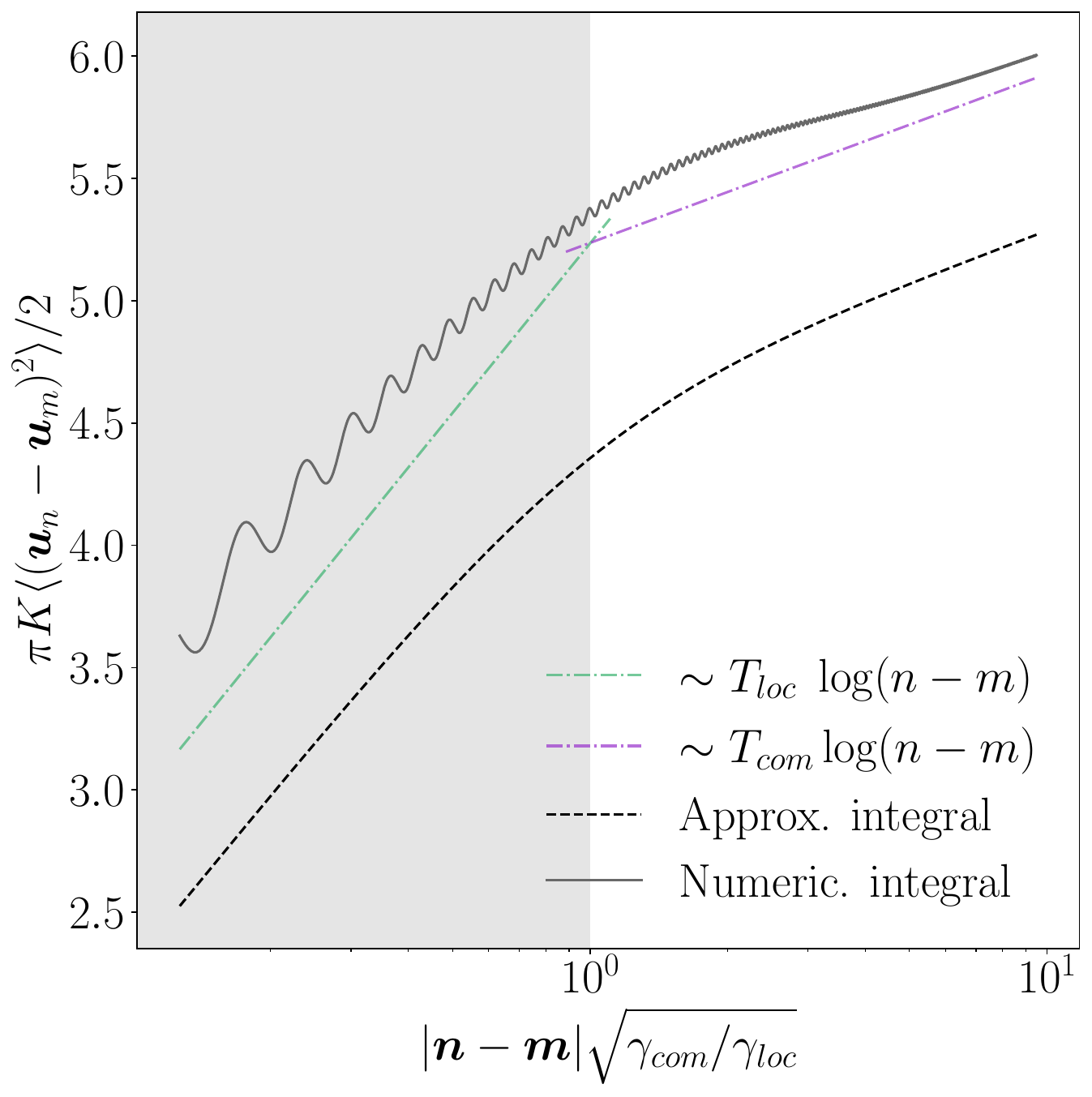}
    \centering
    \caption{Numerical and approximate values of  $\langle (\bm u_n - \bm u_m)^2\rangle$ at $\gamma_{com}/\gamma_{loc} = 10^{-3}$ given by Eqs.~\eqref{eq: res Twotemp correl} and \eqref{eq: res Twotemp correl appendix}. For the numerically solved integral, we use the dispersion relation of a hexagonal lattice.} 
    \label{fig:Fig1}
\end{figure}

To quantitatively describe the emergence of long-range order in our model, we consider the crystalline translational correlation function, defined as \cite{li2019accurate, PhysRevB.3.3939}:
\begin{equation}
    g_{\bm G}(|\bm n - \bm m|) = \left\langle e^{i \bm G \cdot (\bm u_n - \bm u_m)}\right\rangle
                        =  e^{- \langle (\bm G\cdot (\bm u_n - \bm u_m))^2\rangle/2},
    \label{eq: correlation function}
\end{equation}
where $\bm G$ is one of the inner Bragg-peak vectors of the crystal and the last equality follows from the gaussianity of the stationary probability distribution function. 
In a long-range ordered crystal, the correlation function decays to a constant value while in a  quasi-long-range 2D equilibrium crystal
%, due to the quasi-long-range translational order nature of the crystal, 
it exhibits an algebraic decay \cite{mermin1968crystalline, nelson2012bond, PhysRevB.3.3939, https://doi.org/10.1002/cphc.200900755, PhysRevLett.107.155704, PhysRevLett.114.035702}.
%, a condition that goes under the name of quasi-long-range order.
From Eq.~\eqref{eq: correlation function}, it is clear that the asymptotic behavior of the displacement correlation function $\langle (\bm u_n - \bm u_m)^2\rangle$ governs the translational long-range behavior of the system. Specifically, if the displacement correlation function reaches a constant or decays at large distances, true long-range order is expected, otherwise, only quasi-long-range order or short-range order is found.
In the infinite system size limit and at large distances compared to the lattice spacing it can be approximated as (see Appendix~\ref{sec: computation}) :
\begin{equation}
\begin{split}
        \langle (\bm u_n - \bm u_m)^2\rangle &\simeq \dfrac{a^2}{\pi}\int_{\frac{1}{a|\bm n - \bm m|}}^{\frac{\pi}{a}}\dd kk C_{uu}(|\bm k|)\\
        &\simeq \left\{ 
        \begin{array}{ll}
            \dfrac{2T_{com}}{\pi K}\log(\pi|\bm n - \bm m|)\quad|\bm n - \bm m| \gg \delta^{-1}\\
            \\
            \dfrac{2T_{loc}}{\pi K}\log(\pi|\bm n - \bm m|) )\quad|\bm n - \bm m| \ll \delta^{-1}.
        \end{array}\right.
    \end{split}
    \label{eq: res Twotemp correl}
\end{equation}
where $\omega_k^2 = (a\bm k)^2$ was assumed and $\delta = \sqrt{\gamma_{com}/\gamma_{loc}}$ is the adimensionalized natural inverse length scale of the system. At $T_{com} = T_{loc}$ we recover the equilibrium logarithmic increase with distance, proportional to the temperature \cite{mermin1968crystalline}.  In a non-equilibrium setting with $T_{com}\neq T_{loc}$, the divergence is still present, but with $T_{com}$ as the prefactor at large scales and $T_{loc}$ at small scales (still larger than the lattice spacing).  The numerical "exact" computation of the integral for a hexagonal lattice and its comparison with our approximate formula are shown in Fig.~\ref{fig:Fig1}. As expected, the approximations work well at large distances. At short distances, although the global scaling remains the same, strong oscillations appear on top of the logarithmic scaling since the phonons start to feel the discreteness of the lattice. For large $\delta$, the transition between the two regimes happens far away from the lattice spacing allowing for a large range of values over which both scalings are clear and oscillations small.
From Eq.~\eqref{eq: correlation function} and \eqref{eq: res Twotemp correl} we obtain that the translational correlation function exhibits a double power law behavior:
    
\begin{equation}
    g_{\bm G}(|\bm n - \bm m|) \sim \left\{
    \begin{array}{ll}
            |\bm n - \bm m|^{-|\bm G|^2 T_{com}/(2\pi K)}\quad|\bm n - \bm m| \gg \delta^{-1}\\
            \\
            |\bm n - \bm m|^{-|\bm G|^2 T_{loc}/(2\pi K)}\quad|\bm n - \bm m| \ll \delta^{-1},
        \end{array}\right.
    \label{eq: cor}
\end{equation}
where the angle average of the dot product appearing in Eq.~\eqref{eq: correlation function} was taken into account by assuming an isotropic crystal. At non zero $T_{com}$, the large-scale decay of the correlation function is algebraic indicating quasi-long-range order \cite{halperin1978theory}. Nonetheless, in contrast to the equilibrium picture, the average kinetic energy of the system can be arbitrarily high, the only crucial factor for the translational quasi-long-range behavior of the system at large scale, is the temperature of the COM or equivalently, the temperature of the phonons existing on these scales. We also note that, as expected when the COM is conserved at finite global damping ($T_{com} =0$ and $\gamma_{com}\neq 0$), the system displays genuine long-range order as observed in Ref. \onlinecite{PhysRevLett.131.047101} and the HMW theorem is broken. 

We finally point out that the connection between hyperuniformity and the emergence of long-range order is made explicit by Eqs.~\eqref{eq: res Twotemp correl} and \eqref{eq: structurefactor important big big} which show the relationship between the large distance behavior of the displacement correlation function and the low-$k$ limit of the structure factor. Indeed, for large distances, we find in 2D:
\begin{equation}
    \begin{split}
     \langle (\bm u_n - \bm u_0)^2\rangle &\propto \int_{\frac{1}{a|\bm n|}}^{\frac{\pi}{a}}\mathrm{d}k\, k C_{uu}(|\bm k|)\\
     &\propto \int_{\frac{1}{a|\bm n|}}^{\frac{\pi}{a}}\mathrm{d}k\, \frac{S(|\bm k|)}{k}\\
     &\propto \left\{
    \begin{array}{ll}
        S_0\log(|\bm n|) & \mbox{if } S(k) = S_0 + \mathcal{O}(k) \\
        S_1 & \mbox{if } S(k) = S_1k^\beta + \mathcal{O}(k^{\beta + 1})
       \end{array}
    \right.
    \end{split}
    \label{eq: linkSCuu}
\end{equation}
with $\beta>0$. Moreover, the relation between the structure factor and long-range order is not limited to the particular mechanism of hyperuniformity we present in this work but is general and applies as well for different hyperuniform systems such as the critical point of absorbing phase transition \cite{hexner2015hyperuniformity}.

\subsection{Breakdown of the HMW theorem }

The breakdown of the HMW theorem when $T_{com} = 0$ ultimately stems from the fact that large wavelength phonons are thermalized with a bath at 0 effective temperature; a temperature for which, even at equilibrium, long-range order is expected by the HMW theorem. 
In this sense, the theorem is broken in a way expected from its equilibrium definition while simultaneously being strongly broken since the large-scale SDF at $\bm k\to 0$ is constant and does not diverge at all. This can be traced back to the strong hyperuniformity $S(\bm k) \sim \bm k^2$ of our system. We remark that our rationalization of the emergence of stable long-range order based on large wavelength phonons thermalizing at a vanishing effective temperature can be applied also for the non-equilibrium crystal obtained with the random organization model \cite{PhysRevLett.131.047101}.
%As understood from Eqs.~\eqref{eq: structurefactor important big big} and \eqref{eq: mean}, 2D hyperuniform crystals with an exponent below 2 would still break the HMW theorem while displaying a divergent SDF at arbitrarily large wavelength. 

It is important to note that an exponent 2 for the small $k$ structure factor is not a necessary condition to break the HMW theorem. Indeed, as understood from Eq.~\eqref{eq: linkSCuu}, any hyperuniformity in 2D would sufficiently reduce the divergence of the SDF to allow long-range order. This mechanism parallels bird-flocking dynamics, where alignment among neighbors reduces the infrared divergence of the correlation functions to a scaling of $1/k^a$ with $a < 2$ \cite{toner1995long, toner2005hydrodynamics}, allowing for long-range order.

From a mathematical point of view this breakdown of the HMW theorem in the case $T_{com} = 0$ at finite $\gamma_{com}$ is manifested through the regularization of the infrared divergence of the integral used to compute the displacement correlation function (Eq.~\eqref{eq: mean}) via $\gamma_{com}$:
\begin{equation}
    \begin{split}
    \langle (\bm u_n - \bm u_m)^2\rangle \simeq \dfrac{2a^2}{K\pi}\int_{\frac{1}{a|\bm n -\bm m|}}^\frac{\pi}{a} \dd kk &\left(\dfrac{1}{\omega_k^2}\dfrac{T_{com}}{1 + (\omega_k/\delta)^2}\right. \\
     &\quad\left. + \dfrac{T_{loc}}{\delta^2 + \omega_k^2}\right).
    \end{split}
    \label{eq: regula}
\end{equation}
When $T_{com}=0$, the first term vanishes. In the second term, the phonons exhibit a behavior akin to particles with an effective "mass" $\sqrt{K}\delta = \sqrt{K\gamma_{com}/\gamma_{loc}}$ and an adimensionalized dispersion relation $\Omega_k = \sqrt{\delta^2 + \omega_k^2}$. Indeed, an equilibrium system with Hamiltonian \cite{chaikin_lubensky_1995}: 
\begin{equation}
    \mathcal{H}/K =  \dfrac{a^2}{2}\left(2\Tr\left[\bm{\mathsf u}\cdot \bm{\mathsf u}^T\right] - \Tr\left[\bm{\mathsf u}\right]^2\right) + \dfrac{\delta^2}{2} \bm u^2,
\end{equation}
with $\bm{\mathsf u}$ the infinitesimal strain tensor: $\bm{\mathsf u} = (\bm \nabla \bm u + (\bm \nabla \bm u)^T)2$, would produce a similar equilibrium displacement correlation. This effective mass pins down the particles on their ideal lattice position allowing long-range order. Of course, when $T_{com}\neq 0$, the first term of Eq.~\eqref{eq: regula} becomes significant at small $k$ and induces the logarithmic divergence at large scales discussed above.

It is interesting to note that hyperuniformity of equilibrium crystals is solely achieved through long-range or external interactions \cite{torquato2018hyperuniform}, in contrast with our system and that of others, where hyperuniformity is allowed through the breakdown of the FDT and correlated noise \cite{Ikeda2023CorrelatedNA, ma2023theory, lei2019hydrodynamics, song2021kinetic, medina1989burgers}. This is reminiscent of long-range correlations induced by bulk conservation law and dissipation in self-organized criticality \cite{grinstein1990conservation, garrido1990long, van1999randomly, plati2021long, simha2002hydrodynamic, kundu2016long, spohn1983long, dorfman1994generic, grinstein1991generic, bak1992self}. See Ref. \onlinecite{bonachela2009self} for a recent review. 
%from https://arxiv.org/pdf/2210.10009.pdf:
%In fact, certain long-range, algebraic correlations are known to emerge in non-equilibrium situations when conservation laws are present. This is well studied for diffusive systems in non-equilibrium steady states (NESS) [32–34]: unbalanced thermostats at the system’s boundaries lead to nonzero gradients, and correlations between conserved densities at macroscopic distances decay as (system size)−1. This is due to the breaking of detailed balance at the diffusive scale and determined by viscous coefficients, and may be quantitatively described by fluctuating hydrodynamics and macroscopic fluctuation theory (see, e.g., [33–35]).
%Note aalso this very interesting review: https://arxiv.org/pdf/0905.1799.pdf
%Our problem were already known 30 years ago :)
% It is beyond the scope of this paper to review exhaustively the large body of interesting literature devoted to non-conserving SOC models, some aspects of which remain unsettled. But, let us just underline that the state-of-the-art is, as documented in the next section, that none of the considered non-conserving models is truly critical; they just show “apparent scale-invariance” or “dirty criticality” for some decades
Moreover, COM conservation is not the only route to hyperuniformity and systems with multiplicative \cite{jack2015hyperuniformity, ma2023theory} or time correlated noise can have suppressed long-range fluctuations \cite{kuroda2024longrange, Ikeda2023CorrelatedNA}. Additionally, recent work on particular non-equilibrium collective excitations in active solids, known as entropons \cite{caprini2023entropy, caprini2023entropons}, highlight that on a general basis, active forces only add an additional term to the SDF, linked to entropy production but do not affect its phononic part. Thus, active forces alone cannot tame the $1/k^2$ divergence of the phononic part of the equilibrium SDF and in general, increase the density fluctuations \cite{dey2024enhanced, PhysRevE.104.064605, PhysRevLett.131.108301}. Indeed, active short-ranged systems exhibiting hyperuniformity usually lose this property in the presence of a global thermal equilibrium-like noise \cite{huang2021circular, lei2019nonequilibrium, zhang2022hyperuniform, kuroda2023microscopic}. This supports the idea that non-equilibrium-like colored or multiplicative noises are crucial for hyperuniformity and translational long-range order in 2D crystals featuring short-ranged interactions. Further discussions about entropons in our system can be found in Appendix~\ref{sec: appendix entropons}. 
Note however, that hyperuniformity in passive or active scalar field theories has been found even in the presence of a thermal noise for model B type equations \cite{ma2017random, wilken2023spatial, zheng2023universal, deluca2024hyperuniformity}.

Of course, the analysis performed so far predicts the violation of the HMW theorem only in the singular limit $T_{com}=0$, $\gamma_{com}\neq 0$. However, the results obtained in Sec. \ref{sec: qulo} suggest how to approach this physical condition when $T_{com}\neq 0$. Indeed, the key quantity that controls the strength of the quasi-long-range order (and its possible limit to true long-range) is the exponent of the algebraic decay of the
translational correlation function for $|\bm n - \bm m| \gg \delta^{-1}$ in Eq.~\eqref{eq: cor} which is proportional to the ratio $T_{com}/K$. This means that, 
for a finite $T_{com}$, one can still obtain an arbitrary slow decay by increasing the elastic constant $K$. This is clearly a quite limiting and trivial way to enhance the quasi-long-range order of the system since it consists of a direct increase of the particle interaction strength and would also work at equilibrium.    
However, in the next section, we will see how the phenomenology of a realistic model of a non-equilibrium crystal can be mapped into the one of our model with an effective $K$ which depends on the non-equilibrium properties of the realistic system (i.e. dissipation and driving parameters). This will be the key point of our strategy to enhance quasi-long-range through non-equilibrium effects. 

\section{\label{sec:granular} Hard-disk crystal driven by active collisions}

\subsection{\label{sec: coarse grained} A coarse grained model}

We explore a practical application of our theoretical framework to investigate the breakdown of the HMW theorem and the dependence of (quasi-)long-range order on the noise in realistic non-equilibrium systems. To conduct this investigation, we use the model introduced in Ref. \onlinecite{maire_interplay_nodate} with an additional global noise. This can be thought as an underdamped continuous-time analogue of a random organization model \cite{corte2008random} describing sheared suspensions \cite{pine2005chaos}. In the absence of global noise, these models present an absorbing phase transition, are hyperuniform \cite{torquato2018hyperuniform} at the transition point \cite{hexner2015hyperuniformity, ma2023theory} and, if the COM or momentum is locally conserved by the collisions, in the active state \cite{hexner2017noise, bertrand2019nonlinear, lei2019hydrodynamics}. As will be further discussed in the conclusion, this model has been used to describe the experimental realization of vibrated quasi-2d granular systems \cite{brito2013hydrodynamic, maire_interplay_nodate, plati2024quasi}.

We study this system by performing hybrid event-driven/time-stepped molecular dynamics simulations \cite{smallenburg2022efficient} (see Appendix~\ref{sec: molecular dynamics}) of $N$ active hard disks of diameter $\sigma$ and mass $m$ in a 2D square of size $L$ with periodic boundary conditions. We define an arbitrary unit of time $\hat\tau$.
The (granular) temperature $T$ of the system is defined as its global kinetic energy:

\begin{equation}
    T = \dfrac{1}{2}m \sum_{i = 1}^N \bm{v}_i^2,
    \label{eq: def temp}
\end{equation}
with $\bm{v}_i$ the velocity of the particle $i$.
The disks experience a global white bath during their free flight:
\begin{equation}
    \dfrac{\dd \bm v}{\dd t} = -\gamma_{com} \bm v +\sqrt{2\gamma_{com}T_{com}}\bm\eta(t),
    \label{eq: damping}
\end{equation}
with
\begin{equation}
    \langle{\eta}_i^{\alpha}(t){\eta}_j^{\beta}(t')\rangle = \delta(t - t')\delta_{i, j}\delta^{\alpha, \beta}~~~~~\langle {\eta}_i^\alpha(t)\rangle = 0.
\end{equation}
Upon collision, two disks undergo a momentum-conserving active collision \cite{brito2013hydrodynamic, brey2015hydrodynamics, garzo2018enskog}:

\begin{equation}
    \begin{split}
        \bm v_i'&= \bm v_i + \dfrac{1+\alpha}{2}(\bm v_{ij}\cdot \hat{\bm\sigma}_{ij})\hat{\bm\sigma}_{ij} + \Delta \hat{\bm\sigma}_{ij} \\
        \bm v_j'&= \bm v_j - \dfrac{1+\alpha}{2}(\bm v_{ij}\cdot \hat{\bm\sigma}_{ij})\hat{\bm\sigma}_{ij} - \Delta \hat{\bm\sigma}_{ij},
    \end{split}
    \label{eq: collRule}
\end{equation}
where $0\leq\alpha\leq1$ is the coefficient of restitution, $\Delta > 0$ is a velocity injection term, $\bm v_i'$ the post-collision velocity of particle $i$, $\bm v_i$ its pre-collision velocity, and $\hat{\bm\sigma}_{ij}$ and $\bm v_{ij}$ are respectively the unit vector joining particles $i$ and $j$ and the relative velocity between them. In the limit $\Delta = 0$, the usual collision rule of dissipative hard disks is recovered. This model is similar to the one introduced in Ref. \onlinecite{lei2019hydrodynamics} for liquids.

As evidenced by the energy change during a collision
\begin{multline}
    E' - E = m\Delta^2 + m\alpha\Delta| \bm v_{ij}\cdot \hat{\bm\sigma}_{ij}| - m\dfrac{1-\alpha^2}{4}|\bm  v_{ij}\cdot \hat{\bm\sigma}_{ij}|^2,
    \label{eq: collisional E}
\end{multline}
the parameter $\Delta$ controls the intensity of the energy injection at collision. These collisions will thus play the role of the momentum-conserving bath introduced in Sec. \ref{sec: theory} at temperature $T_{loc}$ and associated effective damping $\gamma_{loc}\omega_k^2$.

Hence, to recapitulate the differences and similarities between the harmonic crystal presented in the theory and the granular hard disks of interest here: both models feel an external global Gaussian bath at temperature $T_{com}$ and damping $\gamma_{com}$, however, the active collisions in the granular system are modeled as an \textit{effective} local bath at temperature $T_{loc}$ and damping $\gamma_{com}\omega_k^2$ in the theory. From a physical standpoint, the effective $T_{loc}$ in the granular system is technically a function of every parameter of our simulation including, $T_{com}$ and $\gamma_{com}$ because the energy change at the collision depends on the relative velocities of the particles colliding, as understood from Eq.~\eqref{eq: collRule}, which are affected by the global bath and the packing fraction. This will not change the physical picture given in this article nonetheless, we should keep it in mind. This peculiarity is further discussed in Appendix~\ref{sec: appendix temperature of the bath}.

When $T_{com} = 0$, the collision rule given by Eq.~\eqref{eq: collRule} locally conserves the momentum and the position of the COM. Moreover, the global damping Eq.~\eqref{eq: damping} damps every modes indistinctly. Hyperuniformity is thus expected \cite{lei2019hydrodynamics} and we recover the system studied in Ref. \onlinecite{maire_interplay_nodate} with an absorbing state at low density (a feature absent in the lattice model because $T_{loc}$ is independent on $T_{com}$ and $\gamma_{com}$). Indeed, at low density or packing fraction $\phi = N\pi \sigma^2/4L^2$, the active collisions cannot compensate for the energy dissipated by the drag during the free flight leading the system to an arrested state. Denser systems attain an active steady state where the energy injection at collision is compensated by the dissipation of the damping. 
\subsection{Hyperuniformity and long-range order}

\begin{figure*}[!t]
    \includegraphics[width=1.95\columnwidth,clip=true]{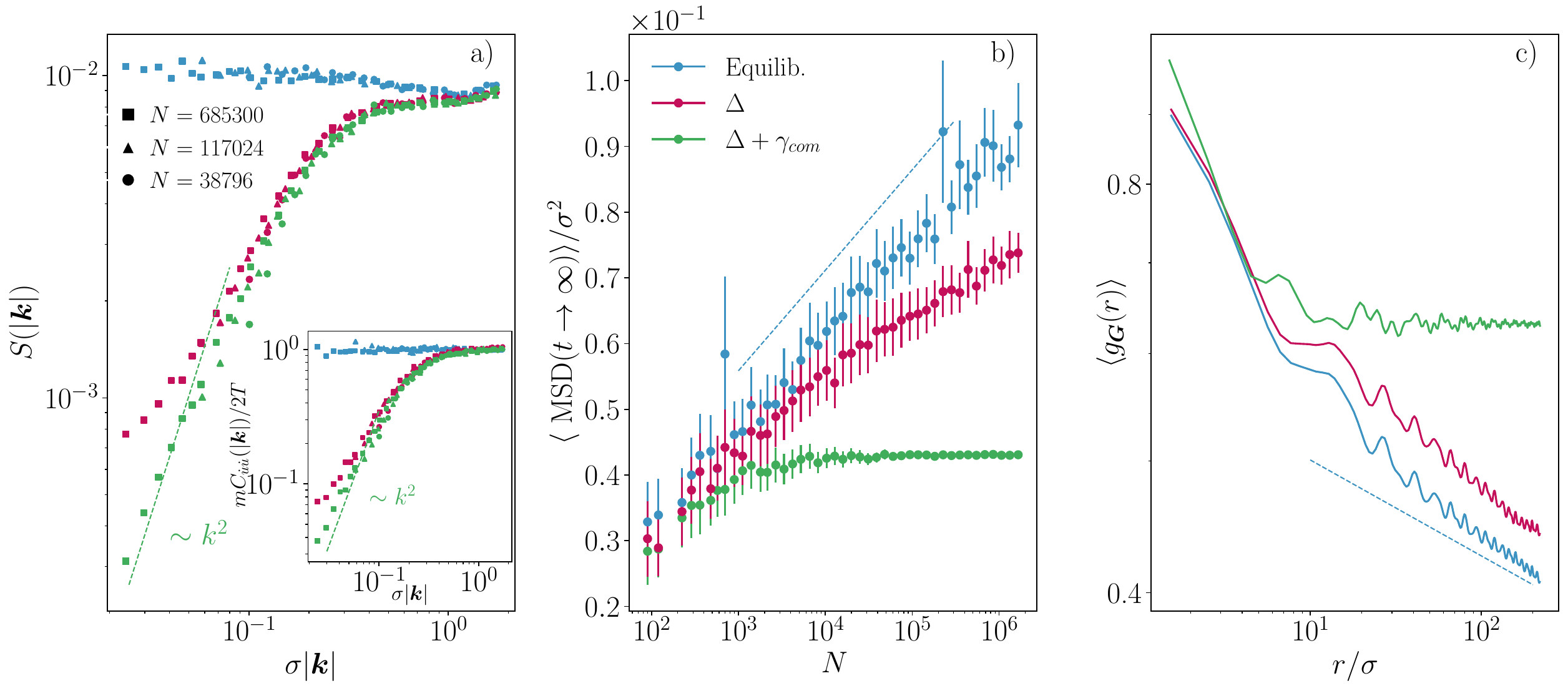}
    \centering
    \caption{Long-range behavior of three different crystalline systems with $T_{com} = 0$. "\textit{Equilib.}" corresponds to pure hard disks ($\alpha = 1$, $\Delta/\sigma \hat\tau^{-1} = 0$ and $\gamma_{com} = 0$), "$\mathit{\Delta}$" corresponds to hard disks with active collisions ($\alpha = 0.95$, $\Delta/\sigma \hat\tau^{-1}  = 0.015$ and $\gamma_{com}/\hat\tau = 0$) and "$\mathit{\Delta + \gamma_{com}}$" corresponds to hard disks with active collisions and global damping ($\alpha = 0.95$, $\Delta/\sigma \hat\tau^{-1}  = 0.015$ and $\gamma_{com}\hat\tau = 0.02$). a) Structure factor of the three systems for three different system sizes in log-log scale. The dashed line represents a power law scaling of $k^{-2}$. The inset is the same plot for the static velocity factor. b) Value of the plateau of the MSD of the three systems as a function of the number of particles, averaged over five different realizations in semi-log. The dashed line is a fit of the logarithmic increase of the equilibrium curve. c) Translational correlation function of the three systems as a function of the distance in log-log with $N = 440538$. The dashed line corresponds to the expected power law scaling of $g_{\bm G}$ extracted from the logarithmic increase of the equilibrium curve in b).} \label{fig:Fig2}
\end{figure*}

To investigate the crystalline order in this system, we simulate crystalline configurations at $\phi = 0.75$ which sets the lattice spacing $a$. This packing fraction corresponds to systems with long-range bond orientational order at equilibrium \cite{bernard2011two} but can give rise to a hexatic or a liquid phase in highly dissipative granular systems \cite{komatsu2015roles}. To avoid these scenarios, we choose a relatively high $\alpha \geq 0.95$ and always check that the crystalline phase is stable. For simulations at $T_{com} = 0$, we stay away from the absorbing region by choosing $\gamma_{com}$ and $\Delta$ such that $\phi$ is above the critical packing fraction of the absorbing to active phase transition and hence, the system has a steady state with finite kinetic energy. Choosing a packing fraction such that $l(\phi)<{\Delta}/{\gamma}$ with $l$ the mean free path, ensures a sufficiently high energy injection at collision which prevents the system to reach the absorbing state  \cite{maire_interplay_nodate}. 

\begin{figure*}[!t]
    \includegraphics[width=1.95\columnwidth,clip=true]{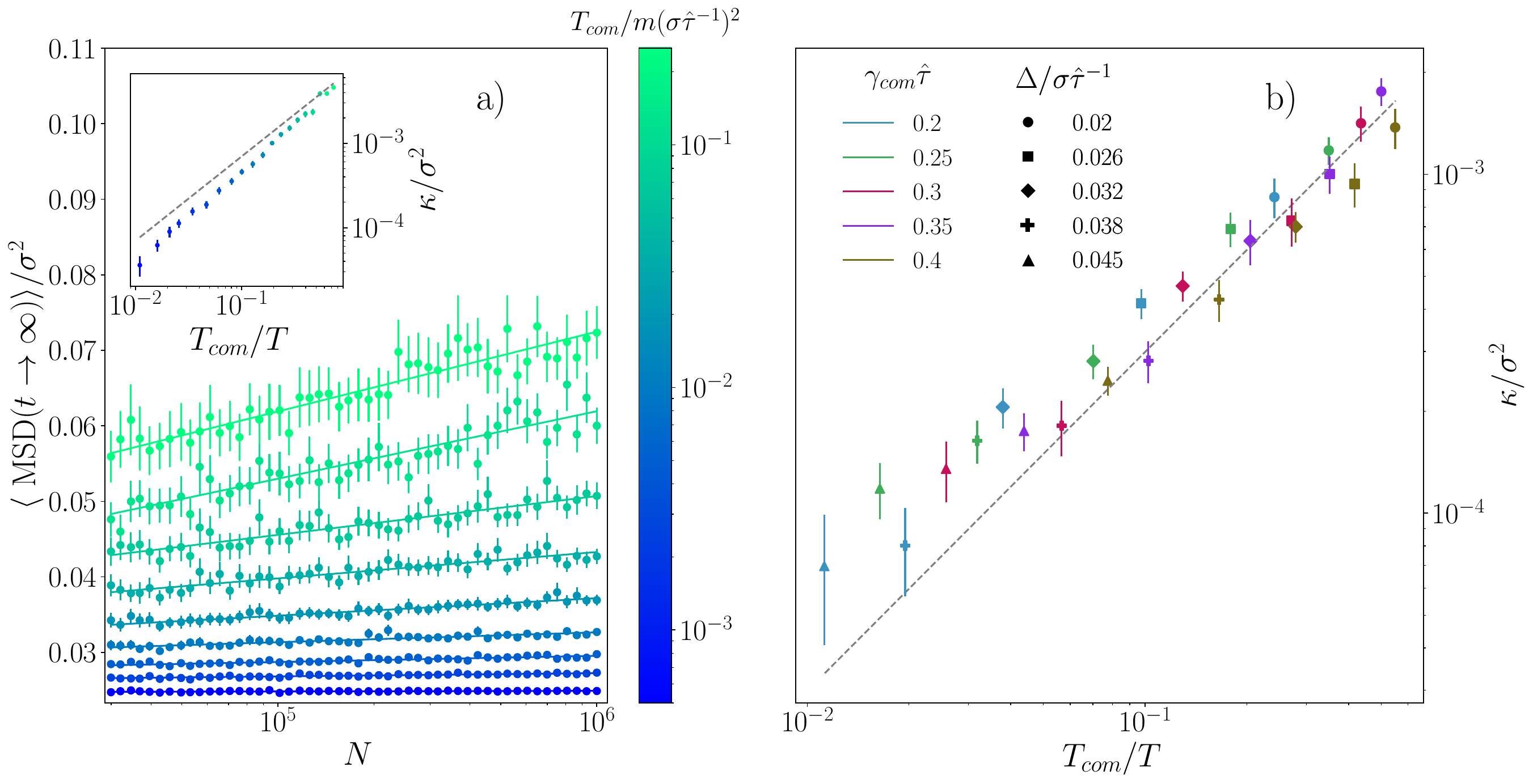}
    \centering
    \caption{Quasi-long-range order dependence on COM temperature. a) Simulation of a system of active hard disks at fixed $\Delta/\sigma \hat\tau^{-1}  = 0.04$, $\gamma_{com}\hat\tau = 0.3$ and $\alpha = 0.95$ for different values of the global Langevin bath temperature $T_{com}$ at fixed $\gamma_{com}$. We assume $T_{loc}$ is fixed and given by Eq.~\eqref{eq: temp}. The main figure shows the dependence on $T_{com}$ (color) of the plateau of the MSD for various system sizes. $T$ is the kinetic energy of the system (see Eq.~\eqref{eq: def temp}). The coefficient in front of the logarithmic increase of the MSD $\kappa$ is shown in the inset as a function of $T_{com}/T$. A linear scaling is given by a grey dashed line as a guide. b) Scaling of the logarithmic increase of the MSD for multiple systems with various $\Delta$ and $\gamma_{com}$ at fixed $T_{com}/m(\sigma \hat\tau^{-1})^2=0.0025$. The underlying finite size analysis on the MSD is performed for 30 systems ranging from $N = 29946$ to $N=1000680$ for each point. A linear scaling is given by a grey dashed line as a guide.} \label{fig:Fig3}
\end{figure*}

We first perform a similar analysis as the one done in Ref. \onlinecite{PhysRevLett.131.047101}, fixing the COM by setting $T_{com} = 0$. The results for three different systems are depicted in Fig.~\ref{fig:Fig2}. Panels a), b) and c) show respectively the structure factor (and the static velocity factor in inset) as a function of $k$, the long-time behavior of the mean square displacement (MSD) for different system sizes and finally the translational correlation function as a function of the distance. The three quantities contain the same information expressed differently. It is illuminating to analyze all three together to better understand the theory given in Sec. \ref{sec: theory} and verify its consistency.  We note that the MSD of a particle in a finite crystal reaches a plateau at large times. The value of this plateau is of course dependent on the size of the cage made by neighboring particles but also on the collective excitations that induce vibrations on the crystal. From the Fourier transform of the displacement $\tilde{\bm u}_{k}(w)$, it can be shown that the correlation $\langle (\bm u_n - \bm u_m)^2\rangle$ for an infinite system corresponds to MSD$(t\rightarrow \infty)$ for a system of size $(a|\bm m - \bm n|)^2$ in the large system size limit. We chose to use the MSD instead of the displacement correlation to check relations derived in the theory part involving the displacement correlation, the former being simpler to measure.

We first focus on the curves "\textit{Equilib}" which correspond to an equilibrium system of pure hard disks ($\alpha = 1$, $\Delta = 0$ and $\gamma_{com} = 0$). As expected, its structure factor is approximately flat (it has in reality an Ornstein-Zernike shape due to the non linear interactions \cite{hansen2013theory}) and reaches a well defined value at small $k$. Its large time MSD grows as a unique logarithm of the particle number as predicted by Eq.~\eqref{eq: res Twotemp correl} and thus its correlation function decreases as a power law with an exponent linked to the logarithmic decrease of the MSD as given by \eqref{eq: cor} and illustrated with the dashed lines. The system displays quasi-long-range order.

We now turn to the \textit{$\Delta$} model, a system with only active collision and no global damping ($\alpha < 1$, $\Delta \neq 0$ but $\gamma_{com} = 0$). 
Hence, the dynamics of the system is ballistic between collisions and the non-equilibrium effects arise purely from the non-equilibrium collision rule Eq.~\eqref{eq: collRule}. The steady state of this system is the result of an interplay between energy injected at collision by $\Delta$ and dissipated by $\alpha$ as given by Eq.~\eqref{eq: collisional E}. In these conditions, the temperature of the system $T$ is a function of $\alpha$ and $\Delta^2$  (see Appendix~\ref{sec: appendix temperature of the bath}, Eq.~\eqref{eq: temp}) and by definition it equilibrates to the local bath temperature $T_{loc}$. Its structure factor has a transient hyperuniform scaling before reaching a plateau at small $k$. The same behavior is found for the static velocity factor. This is incompatible with the modelization of the active collision as an FDT respecting bath at temperature $T_{loc}$ and damping $\gamma_{loc}\omega_k^2$. Indeed, if the active collisions were acting on a coarse grained level like such an equilibrium bath, the structure factor should have resembled the \textit{Equilib} one. It can be shown that, in a hydrodynamic description \cite{brito2013hydrodynamic}, the fast temperature field is highly out-of-equilibrium and acts as a genuinely non-equilibrium bath for the slow velocity field. This effect induces the observed decay. This is however not the main concern of this paper and does not change the phenomenology observed when an additional global damping is added. Hence, the study of this peculiar bath is left for future investigations. The large time MSD of the \textit{$\Delta$} model displays two distinct logarithmic increases corresponding to the two plateaus reached at large and low-$k$ by the structure factor. While the mechanism behind this double logarithmic scaling differs from the one arising due to the difference of temperature between a global and a local COM conserving bath in Eq.~\eqref{eq: res Twotemp correl} (here we recall that there is no global bath since $\gamma_{com} = 0$), the mathematics remain the same, as the low-$k$ structure factor is related to the SDF and consequently the MSD, through Eq.~\eqref{eq: linkSCuu}.  At small wavectors, we conclude that any 2D isotropic harmonic crystal displaying a hyperuniform scaling in $k\in [k_1, k_2]$ and otherwise a constant structure factor, will exhibit this double logarithmic scaling on the displacement correlation; one in the region $r \gg 2\pi/k_1$ and one in $r \ll 2\pi/k_2$ with a transition region in between. $S(k\rightarrow 0)$ and $S(k\rightarrow \infty)$ will respectively be proportional to the prefactor of the long and short range logarithmic increase of the MSD.  If the hyperuniformity extends to $k_1 = 0$ the plateau of the MSD eventually reaches a constant. The translational correlation function of the \textit{$\Delta$} model exhibits as well a double power law decrease, in line with the double logarithmic increase of the plateau of the MSD. The system displays quasi-long-range order as well.

Finally, by introducing a global damping $\gamma_{com} \neq 0$ to the \textit{$\Delta$} model but still with $T_{com} = 0$,  we obtain the "$\mathit{\Delta + \gamma_{com}}$" system which exhibits hyperuniformity on large length scales with a power law consistent with $S(k)\sim k^2$, as predicted in Eq.~\eqref{eq: structurefactor theo}. From the static velocity factor, we confirm that the energy is not evenly distributed across scales but concentrated at fine ones and depleted at larger ones with a $k^2$ power law decrease. From the MSD we observe an initial logarithmic increase followed by a plateau, in line with Eq.~\eqref{eq: cor} when $T_{com} = 0$. This implies the reaching of a constant value for the correlation function at large distances, which is the hallmark of translational long-range order and the breakdown of the HMW theorem, as expected when the COM is conserved in the presence of global damping.

\subsection{Arbitrary enhancement of quasi-long-range order via non-equilibrium effects}

We now turn to the analysis of our full system with $T_{com}\neq 0$ and will show that the relevant temperature is, as expected from the theory, the temperature of the COM and not the temperature of the system $T$ or the temperature of the active collisions $T_{loc}$. Results for a range of temperatures $T_{com}$ of the global bath are presented in Fig.~\ref{fig:Fig3}a). In the main figure, we show various curves of the plateau of the MSD versus the number of particles in the system $N$ for different $T_{com}$ in a semi-log scale. We also provide results for the structure factor and translational pair correlation function in the presence of global bath in Appendix~\ref{sec: structure factor and correlation with global bath}. We ensure $N$ is sufficiently large to only observe the MSD scaling associated with the COM temperature. As expected from Eq.~\eqref{eq: res Twotemp correl}, the increase of the MSD is logarithmic and is greater the larger is $T_{com}$. Of course, we recover a true long-range order in the limit $T_{com}\rightarrow 0$ with a flat plateau of the MSD. In the inset, we extract the coefficient $\kappa$ in front of the logarithmic increase of the MSD as a function of $T_{com}$. That is:
\begin{equation}
    \textrm{MSD}_{T_{com}}(t\rightarrow\infty) = \kappa(T_{com}) \log(N).
\end{equation}
This quantity could have equivalently been obtained from the exponent of the correlation function or from $S(0)$.

For a harmonic crystal, $\kappa$ corresponds, up to a constant factor, to the exponent of the power law decay of the translational correlation function (see Eq.~\eqref{eq: res Twotemp correl} and \eqref{eq: cor}) and is given by
\begin{equation}
    \kappa\sim T_{com}/K.
    \label{eq: kappa}
\end{equation}
However, this simple linear increase of $\kappa$ with $T_{com}$ holds only for harmonic crystals, not for a crystal of active hard disks. The elastic modulus of an equilibrium hard-disk crystal scales trivially with the temperature of the system $T$ \cite{runge1987monte, frenkel1987elastic, laird1992weighted}:
\begin{equation}
    K(T) = \bar K T.
    \label{eq: scaling constant}
\end{equation}
 This implies that the scaling of the MSD of an equilibrium hard-disk system is not expected to grow with the temperature since $T_{com}=T$. While our system is a non-equilibrium hard-disk solid, we make the assumption that Eq.~\eqref{eq: scaling constant} is still a good approximation for the effective elastic constant, thus predicting the following scaling:
 \begin{equation}
     \kappa \sim \dfrac{T_{com}}{T}.
 \end{equation}
This scaling is well verified in our case, as observed in the inset of Fig.~\ref{fig:Fig3} where the dashed line represents a linear increase of $\kappa$ with $T_{com}/T$. 
%We consistently checked that $\kappa$ linearly increases with $T_{com}$ when $T_{com}\ll T_{loc}$ (not shown).

In panel b), we performed the same simulations as for panel a), except that we varied $\Delta$ and $\gamma_{com}$ at fixed $T_{com}$. We observe again approximately a linear scaling of $\kappa$ as a function of $T_{com}/T$ implying that the effective constant of the lattice was roughly constant.

These two analyses indicate that as predicted from the theory, for active hard sphere crystals the relevant parameter to manipulate in order to slow down the decay of the translational quasi-long-range is $T_{com}/T$. Counterintuitively, this implies that increasing the kinetic energy of the system at a fixed center of mass temperature increases the quasi-long-range order. Indeed, we can obtain the same enhancement of quasi-long-range order given by the reduction of $T_{com}$ by keeping it fixed while varying the driving and dissipative parameters $\Delta$ and $\gamma_{com}$.

\section{Conclusions and perspectives towards realistic  systems}\label{sec: expe}

In this paper, we studied the Hohenberg-Mermin-Wagner theorem's applicability to non-equilibrium systems. By examining a theoretical model of a harmonic crystal coupled to a local and a global thermal-like bath, we find that translational long-range order is controlled by the temperature of the center of mass, rather than the overall kinetic energy, as in equilibrium systems. This result also offers a way to better rationalize the violation of the HMW theorem observed in the random organization model \cite{PhysRevLett.131.047101} via a $k$-dependent temperature. Moreover, through numerical simulations of a hard-disk solid driven out-of-equilibrium by active collisions we showed the possibility to suppress density fluctuations and enhance quasi-long-range translational order without neglecting or fine-tuning thermal fluctuations. These insights provide a valuable theoretical understanding of crystalline phases in non-equilibrium systems. 

To conclude, we outline a possible application of our results to suppress density fluctuations in a realistic granular system. From the analysis provided in the previous sections, it is clear that an equilibrium-like global thermal noise prevents long-range translational order and hyperuniformity in both the harmonic and the hard-disk driven solid. In these systems, only the exponent of the power-law decay of the correlation function can be tuned. However, true hyperuniformity or equivalently long-range order are unachievable.
One might be tempted to think that athermal systems, where the effects of thermal fluctuations are negligible, could provide a suitable playground for the actual realization of a two-dimensional long-range ordered crystal. However, things are not so simple. Indeed, once global thermal fluctuations are neglected, it remains practically very difficult to realize a COM-preserving athermal driving mechanism. Moreover, the effect of any confinement will inevitably influence the COM dynamics.
Once this intrinsic difficulty in realizing true long-range order is accepted, one could still hope to find a way to arbitrarily enhance the quasi-long-range order of a system by tuning some experimental parameters following a methodology similar to that used for the simulation of the coarse-grained model in Sec. \ref{sec:granular}.

Quasi-2D vibrated granular fluids \cite{reyes2008effect, PhysRevLett.81.4369, PhysRevE.59.5855, PhysRevE.70.050301, neel2014dynamics, PhysRevLett.89.044301, clerc2008liquid,castillo2019hyperuniform, maire_interplay_nodate} represent realization of the coarse grained model discussed in section \ref{sec:granular} . 
In these systems, macroscopic grains are confined in a box of size $L\times L\times h$ with $h\ll L$ the height of the box. The box is sinusoidally vibrated which imparts vertical energy to the grains. Collisions between particles transfer this vertical energy to the $xy$ components of the velocity. Moreover, tangential friction resulting from collisions with the top or bottom plates slows down the beads in the $xy$ plane. When viewed purely in the $xy$ plane, this system behaves similarly to the coarse-grained model discussed in section \ref{sec:granular}. Indeed, collisions between particles inject energy into the $xy$ plane and play the same role as the $\Delta$ in Eq.~\eqref{eq: collRule} while the damping $\gamma_{com}$ during the free flight in the coarse-grained model is equivalent to the energy loss due to collisions with the roof and the bottom plate in the quasi-2D system. Moreover, the roughness of the two horizontal confining plates implies that each grain-plate collision introduces some randomness to the velocities (without even considering the angular momentum of the grains). Remarkably, this effect has been found to act as a small homogeneous thermal noise \cite{puglisi2012structure} which can be mapped into an effective global temperature $T_{com}$.
The fact that this realistic setup is correctly described by the coarse grain model suggests the possibility of enhancing the quasi-long-range order by tuning the experimental parameters (driving amplitude, frequency, material properties, \dots) to reduce as much as possible the ratio between the effective $T_{com}$ caused by the plate roughness and the horizontal kinetic energy of the system $T$.  Of course, the specific way in which the experimental parameters influence $T_{com}$ and $T$ can be very complex, but after some preliminary analysis, one might in principle be able to find the right combination of them to vary in order to arbitrarily reduce $T_{com}/T$. We look forward to applying these ideas to realistic granular quasi-2D systems in future studies.

\begin{acknowledgments}
We are grateful to Giuseppe Foffi and Frank Smallenburg for carefully reading and commenting on the manuscript. We also thank Andrea Puglisi and Lorenzo Caprini for fruitful scientific discussions. 
This work has been done with the support of Investissements d'Avenir of LabEx PALM (Grant No. ANR-10-LABX-0039-PALM). 

\end{acknowledgments}

\appendix

\section{Discrete real space conserving noises for vectorial quantities}
\label{sec: AppendixNoise}

In Sec. \ref{sec: theory}, we introduced a momentum conserving noise. On a general lattice, such noise can be written as:
\begin{equation}
    \begin{split}
        m\ddot{\bm{u}}_{n} = -\gamma_{loc}\Lambda (\dot{\bm u}_n)  + \sqrt{2\gamma_{loc} T_{loc}} (\bm\nabla\cdot\bm \Xi_n).
    \end{split}
    \label{eq: lattice2}
\end{equation}
The first term is a discrete Laplacian and the second is a discrete divergence of a rank 2 random tensor. Once the discrete differential operators are well defined, this equation can be understood as a discrete analogue of the model B field theory:
\begin{equation}
      m\ddot{\bm{u}}_{n} = -\gamma_{loc}\bm \nabla \cdot (\bm \nabla \bm \dot u_n + \sqrt{2\gamma_{loc} T_{loc}} \bm \Xi_{n}),
\end{equation}
with $\bm\nabla$ a discrete gradient operator. Under this form, $\bm\Xi$ is immediately interpreted as a flux of momentum.
Note however that the global conservation of momentum depends on boundary conditions.

On a general basis, any lattice can be understood as a graph $\mathcal{G}$ with a set of vertices $\mathcal{V}$ and edges $\mathcal{E}$ connecting them. The usual notion of continuous field living on $\mathbb{R}^d$ (or more generally on a manifold) is replaced by the notion of field living on vertices and mapping vertex to vectors. For example in our case $\bm u: \mathcal{V}\to \mathbb{R}^2$. The complexity of dealing with discrete calculus on lattices lies in the fact that the discrete gradient maps fields living on vertices to fields living on the edges while the discrete divergence maps fields living on the edges to fields living on the vertices \cite{lovasz2010discrete, gross2018graph}. Hence, the Laplacian being the composition of the divergence with the gradient maps fields leaving on the vertices to fields living on the vertices. This immediately implies that the divergence in Eq.~\eqref{eq: lattice2} is ill defined if the random tensor $\bm\Xi$ lives on the vertices and not on the edges. Following Ref. \onlinecite{Cavagna2023from}, we will see that this term can be represented through a finite difference of random vertices field. Moreover, in order for the dynamics to be equilibrium-like, the discretization scheme of the differential operators as well as the variance of the noise must be well chosen in order to respect the FDT.

Indeed if we start from a definition of the derivative as a finite difference of terms. We can write the divergence of a tensor $\bm\nabla \cdot \bm a_n$ as finite differences between neighboring points. Then, the corresponding discrete Laplacian $\Lambda$ takes the form $\Lambda(\bm a_n)=\bm \nabla \cdot (\bm \nabla \bm a_n)$ where $\bm \nabla$ is a gradient defined from the chosen finite difference used for the derivative. With these choices, taking $\bm \Xi$ with unit variance leads to a dynamics that correctly respects the FDT. 

As an example, if we define the derivative on a square lattice as the forward finite difference, then: 

\begin{equation}
        \bm \nabla \cdot \bm \Xi_{i, j} = 
        \begin{pmatrix}
            \Xi_{i+1, j}^{x, x}-\Xi_{i, j}^{x, x}+\Xi_{i, j+1}^{x, y}-\Xi_{i, j}^{x, y}\\
            \Xi_{i+1, j}^{y, x}-\Xi_{i, j}^{y, x}+\Xi_{i, j+1}^{y, y}-\Xi_{i, j}^{y, y}
        \end{pmatrix},
        \label{eq: definition derivative}
\end{equation}
where we replaced the usual lower vectorial index $n$ by 2 indexes $i$ and $j$ denoting the $x$ and $y$ position on the lattice. Then, we can check that if $\bm \Xi$ has unit variance:
\begin{equation}
    \langle \bm\Xi_{i,j}^{\alpha, \mu}(t)\bm\Xi_{k, l}^{\beta, \nu}(t')\rangle =
    \delta(t' - t)\delta^{\alpha,\beta}\delta^{\mu, \nu}\delta_{i, k}\delta_{j, l},
\end{equation}
then:
\begin{equation}
    \begin{split}
    \langle (\bm \nabla \cdot \bm \Xi_{i, j}(t))^{\alpha}(\bm \nabla \cdot \bm \Xi_{k, l}(t'))^{\beta} \rangle = & \delta^{\alpha,\beta}\delta(t-t')(4\delta_{i, k}\delta_{j, l}-\\
    &\delta_{i+1, k}\delta_{j, l} - \delta_{i, k}\delta_{j+1, l}-\\
    &\delta_{i-1, k}\delta_{j, l} - \delta_{i, k}\delta_{j-1, l}).
    \end{split}
\end{equation}
Where we recognize the discrete Laplacian found for simple forward finite differences, a sum over first neighbors. A different discretization of the derivative (Eq.~\eqref{eq: definition derivative}) would have given us a different correlation corresponding to the adequate Laplacian to use in order for the system to respect the FDT. For example, a symmetric representation of the derivative would have given rise to a Laplacian with second neighbors \cite{PhysRevX.7.021007}. Conservative discrete equations in real space have been used in a plethora of domains \cite{cavagna2024discrete, PhysRevB.107.224302, PhysRevX.7.021007, Cavagna2023from, glorioso2022breakdown} and naturally arise when a stochastic partial differential equation is discretized to be solved numerically.

\section{Fourier transform convention \label{sec: appendixFourier}}
We choose the following convention for the semi-discrete Fourier transform:

\begin{equation}
    \begin{split}
    \tilde{f}_k(w) &= \int_{-\infty}^{\infty}\dd t\frac{1}{N}\sum_{\bm n} e^{iwt+ia\bm k\cdot \bm n}f_n(t)\\
    f_n(t) &= \dfrac{1}{(2\pi)^3}\int_{-\infty}^{\infty}dw\dfrac{1}{N}\sum_{\bm k} e^{-iwt-ia\bm k\cdot \bm n}\tilde{f}_k(w)
    \end{split}
    \label{eq: defDiscreteFourier}
\end{equation}
with $\bm k= 2\pi \bm n/L$ and $n^{\alpha} \in -\sqrt{N}/2 + 1, \dots, \sqrt{N}/2$. When the spatial continuous limit is taken, we obtain the following Fourier series representation of the real lattice:
\begin{equation}
    f_n(t) = \dfrac{1}{(2\pi)^3}\int_{-\infty}^{\infty}dw\dfrac{a^2}{(2\pi)^2}\int_{-\pi/a}^{\pi/a}\dd \bm ke^{-iwt-ia\bm k\cdot \bm n}\tilde{f}(\bm k,w).
    \label{eq: defContinuousFourier}
\end{equation}
This follows directly from taking the limit $a \rightarrow 0$ while keeping $L$ fixed (or $L\rightarrow\infty$ at fixed $L/\sqrt{N} = a$) and corresponds to an integration in the first Brillouin zone.

\section{Anharmonic lattices 
%and lack of energy at arbitrarily large length scales.
\label{sec: appendix anha}}
The theory described in Sec. \ref{sec: theory} was developed with respect to an harmonic lattice. While anharmonicity leads to benign effect such as a renormalization of the harmonic spring constant, it also leads to energy transfer between modes which could change the phenomenology found for a harmonic lattice. In this section, we argue that this is not the case.

In an system with anharmonic interaction, each mode would still be driven at temperature $\tilde T_{k}$ as in the harmonic case, however, a transfer of energy between them would be observed. We can take as an example simple polynomial forces in real space (without loss of generality, we will stay in 1D, see Ref. \onlinecite{michel2015theory} for a generalization in higher dimension):
$$F_{\textrm{anharmonic}}= -\sum_{b=2}^\infty \beta_b \sum_{\{ \bar n\}} (u_n- u_{\bar n})|u_n-  u_{\bar n}|^{b-1}, $$
with $\beta_b$ coupling constants and $b$ the power law exponent associated to the anharmonic term.
In Fourier space for the mode $k$, such force is written as \cite{bustamante2019exact}:
\begin{multline}
    {F}_{\textrm{anharmonic}}= -\sum_{b=2}^\infty \beta_b \sum_{k_1, k_2, \dots, k_b} V^{{ k}, {k}_1, {k}_2, \dots, { k}_b}_b \times \\ \tilde{{ u}}_{{ k}_1}\tilde{{ u}}_{{ k}_2} \dots \tilde{{u}}_{{ k}_b} \delta_{{k} + {k}_1 + {k}_2 +\dots+ {k}_b}.
    \label{eq: fourier space 2}
\end{multline}

The Kronecker $\delta$ ensures that the momentum is conserved by the scattering of waves. $\beta_b V_b^{k, k_1, k_2, \dots, k_b}$ is the interaction strength of a scattering due to the anharmonicity with power $b$. Such equations are common in wave turbulence \cite{nazarenko2011wave}. For nearest neighbor interaction in 1D, we can show that:
\begin{equation}
    V_b^{k, k_1, k_2,\dots, k_b}\propto \sin(ka/2)\sin(k_1a/2)\dots\sin(k_ab/2),
    \label{eq: interaction 2}
\end{equation}
with $a$ the lattice spacing. These equations are making clear the fact that, with anharmonic terms, modes interact.

A proof of the absence of long-range order in systems with anharmonic interactions is beyond the scope of this work and would most likely require the full stationary distribution to be known as in the equilibrium case [2]. For example, the steady-state distribution of a three-body system, with each particle driven at a different temperature and interacting through a harmonic potential, is known, but already shows great complexity [68].  However, a qualitative argument can be made, suggesting that the overall picture presented in this section should remain largely unaffected at large length scales even in the presence of anharmonic interactions. 

Since the energy is created at the level of collision $k\simeq 2\pi/a$, the most excited modes will be those around this length scale. The question that remains to be answered is: Can these modes transfer sufficiently quickly energy to the modes at $k\to 0$ so that very large wavelength phonons are excited and can reach a non-vanishing temperature? 
The interaction strength between these two scales is given by $V_b^{k\simeq 0, k_1, \dots, k_m \simeq 2\pi/a, \dots, k_b}$ where at least one of the mode is close to 0 (long wavelength) and one mode is close to $2\pi/a$ (collisional/energy creation length scale). However, from Eq.~\ref{eq: interaction 2}, it is clear that the strength of such process goes to 0 with $k\to 0$. Hence, the rate of energy transfer to modes with very large wavelength is vanishingly small with $k$.  Since the continuous damping effectively dissipates any energy  supplied to these modes on a timescale by $\gamma_{com}$, the phenomenology observed for the harmonic case is expected to hold qualitatively. That is, we expect low-k mode to be roughly at temperature $\tilde T_k$, with $T_{k\to 0} = 0$. Note that the argument holds in larger dimensions where the exact expression $V_b^{\bm k, \bm k_1, \bm k_2, \dots, \bm k_b}$ contains term similar to the dispersion relation for each mode $k$, which vanish when $k\to 0$.

\section{Computation of the displacement correlation function\label{sec: computation}}
The displacement correlation function is computed in the infinite size limit where the discrete Fourier transform over wavevectors is approximated as a continuous Fourier transform (see Appendix~\ref{sec: appendixFourier}):

\begin{equation}
    \begin{split}
        \langle (\bm u_n - \bm u_m)^2\rangle 
        &= \dfrac{a^4}{(2\pi)^4}\int \dd \bm k \dd \bm q (e^{ia\bm k\cdot\bm n} - e^{ia\bm k\cdot \bm m})\\
        &\times (e^{ia\bm q\cdot \bm n} - e^{ia\bm q\cdot \bm m})\langle\tilde {\bm u}(k, t = 0)\tilde {\bm u}(q, t = 0)\rangle  \\
        &= \dfrac{a^2}{2\pi^2}\int\dd \bm k (1 - \cos(a\bm k \cdot  (\bm n - \bm m))C_{u u}(\bm k)
    \label{eq: mean}
    \end{split}
\end{equation}

Since we only are interested in the (quasi-)long-range behavior of our system, we compute the long distance limit of this object. We decompose the integral into two intervals. From $0$ to $1/a|\bm n - \bm m|$, if $T_{com}\neq 0$ the integral is bounded by a term in $1/\omega_k^2$ and can be neglected. For the second interval from $1/a|\bm n - \bm m|$ to $2\pi/a$, the oscillating part averages out. Similar arguments work for $T_{com} = 0$. Up to a constant, at long distances, Eq.~\eqref{eq: mean} is thus approximated as:

\begin{equation}
    \begin{split}
        \langle (\bm u_n - \bm u_m)^2\rangle &\simeq \dfrac{a^2}{\pi}\int_{\frac{1}{a|\bm n - \bm m|}}^{\frac{\pi}{a}}\dd k C_{uu}(k)k\\
        &\simeq \left[\vphantom{2 T_{com} \log (\pi  |n - m|)}(T-T_{com})\log \dfrac{\gamma_{com}/\gamma + \pi^2 }{\gamma_{com}/\gamma + 1/|\bm n - \bm m|^2}\right.\\
        &~~~\left.+~2 T_{com} \log (\pi  |\bm n - \bm m|)\vphantom{(T-T_{com})\log \dfrac{\pi^2 }{\gamma_{com}/\gamma + 1/|\bm n - \bm m|^2}}\right]/\pi K\\
        &\simeq \left\{ 
        \begin{array}{ll}
            \dfrac{2T_{com}}{\pi K}\log(\pi|\bm n - \bm m|)\quad|\bm n - \bm m| \gg \delta^{-1}\\
            \\
            \dfrac{2T_{loc}}{\pi K}\log(\pi|\bm n - \bm m|) )\quad|\bm n - \bm m| \ll \delta^{-1}.
        \end{array}\right.
    \end{split}
    \label{eq: res Twotemp correl appendix}
\end{equation}
which gives Eq.~\eqref{eq: res Twotemp correl}.

\section{Entropons in a two temperature delta correlated bath \label{sec: appendix entropons}}

Entropons, as discussed in Caprini \textit{et al.}'s recent work  \cite{caprini2023entropy, caprini2023entropons} are general non-equilibrium collective excitations linked to the spectral entropy production and are a consequence of a Harada-Sasa like relation \cite{harada2005equality, nardini2017entropy, caprini2023entropons}. 
Indeed, for a generalized Langevin equation:
\begin{equation}
        \ddot{\tilde{{u}}}_{k}(t) = - K\omega_k^2\tilde{{u}}_{k}- \int_{-\infty}^{\infty} \dd t' \Gamma(k, t -t')\dot{\tilde{u}}_{k}(t') + F + \tilde{{\mathcal{E}}}_k(t),
    \label{eq: Delta model 1}
\end{equation}
with $\Gamma(t)$ the friction kernel and $\langle\tilde{\mathcal{E}}_k(t')\tilde{\mathcal{E}}_q(t)\rangle=\tilde\nu(k, t - t')\delta_{k, -q}$, we can prove that, if the force $F$ is even under time reversal the following relation holds \cite{crisanti2012nonequilibrium}:

\begin{equation}
    \dfrac{C_{uu}(k, w)}{\nu( k, w)}=\dfrac{\Im(R( k,w))}{w\Re(\Gamma( k,w))}+\dfrac{\sigma^g( k, w)}{2\Re(\Gamma( k,w))^2w^2},
\end{equation}
where
\begin{equation}
    \sigma^g( k, w) = 2\Im\langle \tilde u( k, w)F(- k,-w)\rangle\Re(\Gamma( k,w))\nu^{-1}( k,w)w
\end{equation}
is the entropy production of the generalized Langevin equation, $R(\mathbf k, w)$ is the response of the displacement to an instantaneous force while $\Re$ and $\Im$ are the real and imaginary part of the function in front of it. Note that this formula holds independently on the choice of the noise correlation and damping. Notably in the case we are interested, $\Gamma( k, w) = \gamma_{com} + \gamma_{loc}\omega_k^2$ and $\nu(w)/\Gamma( k, w)=2\tilde T_k$, without additional forces, the spectral entropy production $\sigma^g$ is equal to 0 and the non-equilibrium character of the system is only included in the effective temperature $T_{\text{eff}}( k, w) = \nu( k, w)/(2\Gamma( k, w))$. In this sense, we might consider that entropons are absent and only phonons at an effective temperature exist in our system since the dynamic SDF reads:
\begin{equation}
    C_{uu}( k, w)= \dfrac{2\tilde T_k}{w}\Im(R( k, w)).
    \label{eq: no entropons}
\end{equation}

The entropy production equal to 0 might come as a surprise since it is clear that as long as $T_{com}\neq T_{loc}$, heat flows from one reservoir to another and thus entropy is continuously created. However, the particle itself is oblivious to these heat transfers between baths because they collectively act as a unique equilibrium-like reservoir with an effective temperature and viscosity \cite{lee2018stochastic, murashita2016overdamped, van2010three}.  From the point of view of the particle, the system is thus too coarse-grained to probe these non-equilibrium effects and heat fluxes between reservoirs are invisible to dynamical observables \cite{puglisi2006dynamics, sarracino2010granular}. 

To circumvent this issue, we can rewrite our model in an equivalent form for the particle, with an auxiliary variable \cite{plati2023thermodynamic, Plati2020, glatt2020generalized, pavliotis2016stochastic, goychuk2009viscoelastic, kupferman2004fractional, dygas1986singular,puglisi2009irreversible, sarracino2010granular}. This auxiliary variable provides enough information about the system to probe the heat fluxes between reservoirs and compute the spectral heat flow between them. The heat flux for Eq.~\eqref{eq: FinalFourier} is given by \cite{plati2023thermodynamic}:

\begin{equation}
    m\dot Q( k, w) = \dfrac{4\gamma_{com}\gamma_{loc}\omega_k^2(T_{com}-T_{loc})w^2}{(K\omega^2_k - mw^2)^2 + (\gamma_{com}+\gamma_{loc}\omega_k^2)^2w^2}.
    \label{eq: spectral heat prod}
\end{equation}
From which we can prove in our system without additional forces that:
\begin{equation}
    C_{uu}( k, w)= \dfrac{2T_{loc}}{w}\Im(R( k, w))+\dfrac{1}{2w^2\gamma_{loc}\omega_k^2}\dot Q( k, w),
    \label{eq: calorons2}
\end{equation}
or
\begin{equation}
    C_{uu}( k, w)= \dfrac{2T_{com}}{w}\Im(R( k, w))-\dfrac{1}{2w^2\gamma_{com}}\dot Q( k, w).
    \label{eq: calorons}
\end{equation}
The first term represents the usual equilibrium phonons at temperature $T_{loc}$ while the second term is clearly a non-equilibrium term, vanishing when $T_{com}=T_{loc}$ and proportional to the heat transferred between reservoir. This new term, arising from a rewriting of the effective temperature is very similar to entropons except that it is better written as proportional to the heat transfer instead of the entropy production due to Fourier's law. Note however that, contrary to entropons arising from active forces, this new term might be negative according to which bath we chose as the one with the reference temperature in front of the linear response function. 

The equivalence between Eqs.~\eqref{eq: calorons2}-\eqref{eq: calorons} and Eq.~\eqref{eq: no entropons} can be proved more generally with multiple exponentially correlated reservoirs and external non-conservative forces using the markovian auxiliary variable representation of the generalized Langevin equation. Hence, for non-interacting systems, the effective temperature in $\bm k, w$ space, obtained from the ratio between the autocorrelation and the response function can be expressed as the temperature of a bath plus terms proportional to the heat flow between all the other baths. We look forward to developing these ideas in future work.

\section{Hybrid event-driven/time-stepped molecular dynamics}
\label{sec: molecular dynamics}

Since the potential of hard-disks particles is discontinuous, usual time-stepped molecular dynamics methods are not suitable for simulations of such systems. Instead, we use event-driven methods \cite{smallenburg2022efficient} where the time before the collision of two particles $i$ and $j$ denoted $t^{col}_{ij}$ can be analytically computed:

\begin{equation}
    |\bm r_i(t^{col}_{i j}) + \bm r_j(t^{col}_{i j})| = \dfrac{\sigma_i + \sigma_j}{2},
\end{equation}
with $r_i(t)$ the position of particle $i$ at time $t$. 
For particles free flying or undergoing constant viscous drag, $t_{col}^{ij}$ can be found exactly from the initial velocities and positions of the particles.

For viscous friction, the time before the next collision is:
\begin{equation}
    t^{col}_{ij} = -\log\left(1-\gamma_{com}\delta t_{ij}\right)/\gamma_{com},
\end{equation}
with:
\begin{equation}
    \delta t_{ij}=\frac{-b - \sqrt{b^2 - \bm v_{ij}^2(\bm r_{ij}^2-(\sigma_i+\sigma_j)^2/4)}}{\bm v_{ij}^2}
\end{equation}
where $b = \bm r_{ij}\cdot \bm v_{ij}$ and $\bm r_{ij}$ and $\bm v_{ij}$ are respectively the relative position and velocity of particles $i$ and $j$ at the moment the subsequent collision time is computed.
From every collision time and the collision rule \eqref{eq: collRule} the full dynamics of the system can be solved exactly in the absence of $T_{com}$. When the bath has a non-zero temperature, we must include the effect of the thermostat in the system by adding events every $\delta t_{noise}$ where each particle is kicked by an amount $\sqrt{2\gamma_{com}T_{com}\delta t_{noise}}/m$.
The FDT is respected for every value of $\delta t_{noise}$ because of the continuous nature of the damping \cite{ma2017fluctuation}. We nonetheless chose $\delta t_{noise} \ll\tau_f \equiv 1/\omega(\phi, T)$ the mean free time.

    \begin{figure*}[!ht]
        \includegraphics[width=0.95\textwidth]{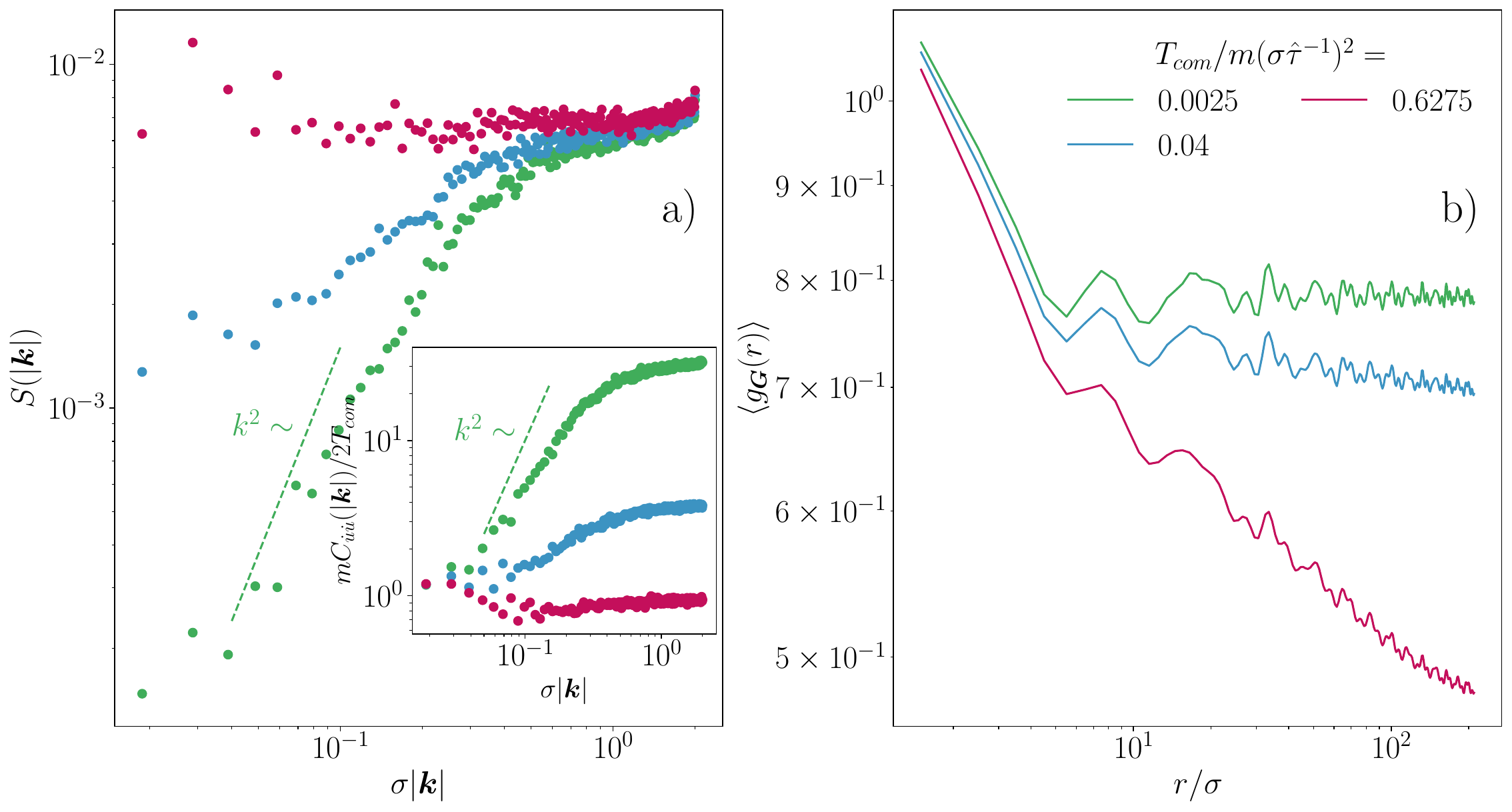}
        \centering
        \caption{Effect of a global bath on the $\Delta$+$\gamma$ model for the structure factor and the translational pair correlation function. a) Main figure: structure factor as a function of the wavenumber. For lower $T_{com}$ it is easier to observe a transient hyperuniform scaling $k^2$. In the limit $T_{com} = 0$, the hyperuniform scaling continues to $k=0$. Inset: static velocity factor divided by $T_{com}$ as a function of $k$. We obtain the limit $mC_{\dot{u}\dot{u}}(|\boldsymbol{k}|\to 0)/2=T_{com}$ as expected from $T_{k\to 0}=T_{com}$. b) Translational correlation function as a function of $r$. Increasing  $T_{com}$, we observe a stronger quasi-long-range power law decay. In the limit $T_{com}=0$, $g_{\boldsymbol G}$ flattens and true long-range order is obtained. $N = 500080$, $\phi = 0.765$, $\Delta/\sigma \hat\tau^{-1}  = 0.04$, $\gamma_{com}\hat\tau = 0.3$ and $\alpha = 0.95$.} 
        \label{fig:Fig appendix}
    \end{figure*}

\section{Temperature of the active collision bath \label{sec: appendix temperature of the bath}}

For the model including only active collisions ($\gamma_{com} = 0)$, the average energy change during a collision can be obtained from the collision rule \eqref{eq: collisional E}:
\begin{multline}
    \langle E' - E\rangle_{coll} = m\Delta^2 + m\alpha\Delta\langle| \bm v_{ij}\cdot \hat{\bm\sigma}_{ij}|\rangle_{coll}\\ - m\dfrac{1-\alpha^2}{4}\langle|\bm  v_{ij}\cdot \hat{\bm\sigma}_{ij}|^2\rangle_{coll},
    \label{eq: collisional E averaged}
\end{multline}
where the average can be computed from the assumption of molecular chaos, gaussianity and using the cross section of hard disks \cite{pagonabarraga2001randomly, brito2013hydrodynamic, maire_interplay_nodate}. In the steady state, on average collisions do not inject or retrieve energy from the system. The steady-state temperature $T_{loc}$ is thus given by the zero of Eq.~\eqref{eq: collisional E averaged} which is \cite{brito2013hydrodynamic}:
\begin{equation}
    T_{loc} = \Delta^2\left(\dfrac{\epsilon  + \sqrt{\epsilon^2 +4m(1-\alpha^2)}}{2(1-\alpha^2)}\right) ^2
    \label{eq: temp}
\end{equation}
with $\epsilon = \alpha\sqrt{\pi m}$. This can be taken as the temperature $T_{loc}$ of the local bath.
It can be shown \cite{maire_interplay_nodate} that when damping is added to the system, far from the absorbing state, the new temperature of the system $T$ keeps the exact same form as Eq.~\eqref{eq: temp} but with $\epsilon = \alpha\sqrt{\pi m} - \gamma_{com}\sigma\sqrt{\pi m}/2\Delta\phi\chi(\phi)$ with $\chi(\phi)$ the radial distribution function at contact.

This implies that the temperature of the system $T$ does not behave as the temperature of a system subject to a Langevin bath at temperature $T_{loc}$ and damping $\gamma_{loc}\omega_k^2$ on top of which we add an additional damping $\gamma_{com}$. This is due to the dependence of Eq.~\eqref{eq: collisional E averaged} on the relative velocity of particles at collision which indicates that $T_{loc}$ should itself be dependent on the other sources of energy change in the system. It will not cause any significant practical problem but this is a peculiarity that should be kept in mind.

\section{Structure factor and translational correlation function in the presence of a global bath}
\label{sec: structure factor and correlation with global bath}
In Fig.~3 we analyzed the dependence of the MSD on $T_{com}$. As described with Fig.~2 and in the theory, the plateau of the MSD, the low-$k$ structure factor and the long-range behavior of the translational pair correlation function contain the same information as given by Eqs.~\eqref{eq: correlation function} and \eqref{eq: linkSCuu}. Nonetheless, as a consistent check, it is interesting to confirm numerically these predictions in our active hard disk model. In Fig.~\ref{fig:Fig appendix}a), we provide the structure factor of systems driven at different global bath temperatures.  As expected, we observe that the smaller $T_{com}$ is, the lower the structure factor can go, in agreement with Eq.~\eqref{eq: structurefactor theo}. The increase of the plateau of the MSD with system size is proportional to $S(0)$, however, in the main text we use the MSD since we consider it provides better data for finite system size (at infinite system size it would be easier to measure the plateau of the structure factor at low-$k$). In the inset, we provide as well, the velocity static factor that converges, as it should for small $k$ to $T_{k\to 0}=T_{com}$. In Fig.~\ref{fig:Fig appendix}b), we give the translational pair correlation for these systems. Their long-range behavior is in agreement with Eq.~\eqref{eq: cor}: they exhibit a power law decay with an exponent increasing with $T_{com}$. The exponent of the power law decay is proportional to the increase of the MSD plateau, however, as for the structure factor, data are cleaner with the MSD.

\bibliographystyle{unsrt}
\bibliography{bib}

\end{document}